\documentclass[11pt]{article}
\usepackage{amsmath,amssymb,color,graphics,epsfig,tikz}

\usepackage{amsfonts}

\newcommand{\be}{\begin{equation}}
\newcommand{\ee}{\end{equation}}
\newcommand{\bea}{\setlength\arraycolsep{2pt} \begin{eqnarray}}
\newcommand{\eea}{\end{eqnarray}}
\newcommand{\nn}{\nonumber}

\def\fft#1#2{{\frac{#1}{#2}}}

\def\0{{\sst{(0)}}}
\def\1{{\sst{(1)}}}
\def\2{{\sst{(2)}}}
\def\3{{\sst{(3)}}}
\def\4{{\sst{(4)}}}
\def\5{{\sst{(5)}}}
\def\6{{\sst{(6)}}}
\def\7{{\sst{(7)}}}
\def\8{{\sst{(8)}}}
\def\sst#1{{\scriptscriptstyle #1}}

\begin{document}

\begin{flushright}
\end{flushright}

\vspace{25pt}
\begin{center}
{\large {\bf Understanding the analysis of wave function}}

\vspace{10pt}
Wei Yang

\vspace{10pt}

{\it College of Physics and Electronic Information Engineering, Guilin University of
Technology, Guilin, 541004, China}

\vspace{40pt}

\underline{ABSTRACT}
\end{center}

We obtained the order $\rho$ and the type $\sigma$ of wave function with power-law potentials, and found that the order and the type are compatible with the condition $|\sigma|\rho =\sqrt{v_m}$ except to $m=-2$. At the same time, we ansatz that the wave function satisfaction relation $\psi(r)=f(r)\exp[g(r)]$, where $f(r),g(r)$ are polynomials, and the power of $g(r)$ is no more than the order $\rho$.

\vfill {\footnotesize Emails: weiyang@glut.edu.cn}
\thispagestyle{empty}

\pagebreak



\newpage

\section{Introduction}
\label{Sec.introduce}
The exact solutions to the Schr\"{o}dinger equation play crucial roles in physics. It is well-known that there are only several potentials can be exactly solved, while most of them require some approximation methods to obtain the solutions. The problem of inverse-power potential $1/r^n$ has been widely emerge in classic mechanics and quantum mechanics. For instance, the interatomic interaction potential in molecular physics, one-electron atoms, and hadronic and Rydberg atoms so on, require considering the inverse-power potentials~\cite{Bransden}.

It is well known that the general framework of the non-relativistic quantum mechanics is well understood~\cite{Isham,Dirac}, whose predictions have been confirmed by careful observations~\cite{Born}.
It is important to know whether some familiar problems have a more general scheme. On this purpose, it is worthwhile to solve the Schr\"{o}dinger equation with general methods.
Recently, the study of higher order anharmonic potentials have been much more desirable to physicists~\cite{Dong:1999ns,Dong:1998sx}, who want to understand the relate between the wave function and the potential. The inverse-power potential $V(r)=A/r+B/r^2+C/r^3+D/r^4$ has been discussed by S.~\"{O}zcelik \textit{et al.}~\cite{Simsek}, who gave a solution under the coefficients of potential must be satisfy some restriction conditions.

The purpose of this paper will give the concepts of the order and the type in the wave function, and use this concepts to provide a general method to solve the Schr\"{o}dinger equation.

The work is organized as follows. In Sec.~ \ref{Sec.formalism}, we will prove the order $\rho$ and the type $\sigma$ of the wave function satisfying $|\sigma|\rho =\sqrt{v_m}$, except to $m=-2$. In Sec.~\ref{Sec.model1},~\ref{Sec.model2}, we use this method to calculate the wave function and the energy of harmonic oscillator and Coulomb potential. In Sec.~\ref{Sec.conclusion}, we will give some conclusions for this paper.

\section{Schr\"{o}dinger equation}
\label{Sec.formalism}
\subsection{Positive power series potentials}
\label{Sec.formalism1}
Throughout this paper the natural unit $\hbar=1$ and $\mu=1/2$ are employed. First we consider the one-dimensional Schr\"{o}dinger equation with positive power series potentials $V(r)= v_0+v_1r+v_2r^2+v_3r^3+\cdots$. 
\begin{align}
\Psi''(r) = [V(r)-E]\Psi(r)\,.
\end{align}
Where the energy $E=k^2$. Suppose the potential $V(r)$ convergence in region $|r-r_0|<R(R>0)$, we know from differential equation theory that the solution of second order differential equation will convergence in the same region, see the appendix~\ref{Sec.appendix1}. So the wave functions are analysis in the region $|r-r_0|<R$.

To consider the the order $\rho$ and type $\sigma$ of wave function, we need to know the limit of wave function $\psi(r)$ in $r\rightarrow\infty$. For simplicity, we assume that the maximum power term of $V(r)$ is $r^m$. So when $r\rightarrow\infty$, that the Schr\"{o}dinger equation will become
\begin{align}
\psi''(r) = v_m r^m\psi(r)\,.
\label{eq:1}
\end{align}
Where the number $m>0$.  As the wave function is analysed in real axis, we can suppose $\psi(r)$ by series expansion
\begin{align}
\psi(r) =\sum_{n=0}^\infty a_n r^n\,.
\label{eq:2}
\end{align}
Substituting the Eq.~(\ref{eq:2}) into Eq.~(\ref{eq:1}), we arrive at the following equation
\begin{align}
\sum_{n=0}^{m-1} (n+1)(n+2)a_{n+2}r^n+\sum_{n=m}^\infty [(n+1)(n+2)a_{n+2}- a_{n-m}v_m]r^n=0\,.
\end{align}
So, we have the following recurrence relationship
\begin{align}
(n+1)(n+2)a_{n+2}=a_{n-m}v_m \,,\nn\\
a_2=\cdots=a_{m+1}=0 \,.
\end{align}
Further, we can calculate
\begin{align}
\psi(r)=& a_0[1+\sum_{n=1}^\infty \fft{(v_m)^n r^{(m+2)n}}{(m+1)(2m+3)\cdots((m+2)n-1)(m+2)^nn!}]\nn\\
&+a_1[r+\sum_{n=1}^\infty \fft{(v_m)^n r^{(m+2)n+1}}{(m+3)(2m+5)\cdots((m+2)n+1)(m+2)^nn!}]\,.
\end{align}
From the entire function theory, that the order $\rho$ and the type $\sigma$ can be repressed by
\begin{align}
\rho= -\overline{\lim_{n\rightarrow\infty}}\fft{n\ln n}{\ln|a_n|}\,,\nn\\
(\sigma e \rho)^{\fft{1}{\rho}}=\overline{\lim_{n\rightarrow\infty}}n^{\fft{1}{\rho}}|a_n|^{\fft{1}{n}}\,.
\label{eq:3}
\end{align}
Where the over line mean superior limit. From the series expansion of $\psi(x)$ with a litter calculate, we have
\begin{align}
\rho= \fft{m+2}{2}\,,\qquad |\sigma| \rho= \sqrt{v_m}\,.
\label{eq:8}
\end{align}
We have used the Stirling's approximation.
\subsection{Negative power series potentials}
In this we consider the Schr\"{o}dinger equation with negative power series potentials $V(r)= \cdots+v_{-2}r^{-2}+v_{-1}r^{-1}+v_{0}$, then the radial wave function obeys
\begin{align}
\Psi''(r) = [V(r)+\frac{l(l+1)}{r^2}-E]\Psi(r)\,.
\end{align}
We can see this potential have two  singular points, one is the original point, the other is infinity point. We have discussing the positive power series potentials in~ \ref{Sec.formalism1}, in subsection we will stress for the negative power series potentials.
Assume that the minimum power term of $V(r)$ is $r^{-m}$. So when $r\rightarrow 0$ that the Schr\"{o}dinger equation become
\begin{align}
\psi''(r) = v_{-m} r^{-m}\psi(r)\,.
\label{eq:21}
\end{align}
For $m=0$, this equation have solution $\psi(r)=c_1\exp[\sqrt{v_{0}}r]+c_2\exp[-\sqrt{v_{0}}r]$, obviously, the order and type of $\psi(r)$ was gave $\rho=1,\sigma=\pm\sqrt{v_{0}}$, this satisfy the relations $\rho|\sigma|=\sqrt{v_{0}}$ and $\rho =\fft{-m+2}{2}$.

For $0<m\leq 2$, we can know from the appendix~\ref{Sec.appendix2} that Eq.~(\ref{eq:1}) have solution like $r^\alpha\sum_{n=0}^\infty a_n r^n$.(Note for $m=2$, we only consider the case where the modulus of the difference between the two roots of the characteristic equation not equal an integer).

Suppose $\psi(r)$ have series expansion
\begin{align}
\psi(r) =r^\alpha\sum_{n=0}^\infty a_n r^n\,.
\label{eq:22}
\end{align}
Substituting the Eq.~(\ref{eq:22}) into Eq.~(\ref{eq:21}), have the following equation
\begin{align}
\sum_{n=0}^{-m+1} (\alpha+n)(\alpha+n-1)a_{n}r^n+\sum_{n=-m+2}^\infty [(\alpha+n)(\alpha+n-1)a_{n}- a_{n+m-2}v_{-m}]r^n=0\,.
\end{align}
So, we have recurrence relationship
\begin{align}
(\alpha+n)(\alpha+n-1)a_{n}=a_{n+m-2}v_{-m}\,,\nn\\
a_0=\cdots=a_{-m+1}=0 \,.
\label{eq:23}
\end{align}
For $m=1$, we can calculate
\begin{align}
\psi(r)=& a_1r^\alpha[r+\sum_{n=1}^\infty \fft{(v_{-m})^{n-1} r^{n}}{[(\alpha+1)(\alpha+2)\cdots(\alpha+n-1)][(\alpha+2)(\alpha+3)\cdots(\alpha+n)]}]\,.
\end{align}
From the Eq.~(\ref{eq:3}) and using the Stirling's approximation, we have
\begin{align}
\rho= \fft{1}{2}=\fft{-m+2}{2}\,,\qquad |\sigma| \rho= \sqrt{v_{-m}}\,.
\end{align}
For $m=2$, we can see that the Eq.~(\ref{eq:23}) hold only $n$ a special  number and obey $(\alpha+n)(\alpha+n-1)=v_{-m}$, than we have $\psi(r)\propto r^n$.  So the order is give $\rho=0$ and the type $\sigma=n$. Only in this case the relation $|\sigma| \rho= \sqrt{v_{-2}}$ not satisfied.

Next, let's investigate $m>2$, in this case we can change $r=1/t$, it is show that the new variable $t\rightarrow\infty$, when $r\rightarrow 0$, here the Eq.~(\ref{eq:21}) will become
\begin{align}
\fft{d^2\psi(t)}{dt^2}+\fft{2}{t}\fft{d\psi(t)}{dt}-v_{-m}t^{m-4}\psi(t)=0\,.
\end{align}
If we define $\chi(t)=t\psi(t)$, than
\begin{align}
\fft{d^2\chi(t)}{dt^2}-v_{-m}t^{m-4}\chi(t)=0\,.
\label{eq:25}
\end{align}
From the appendix~\ref{Sec.appendix2}, we know when $m>2$ the Eq.~(\ref{eq:25}) also have series solution $\chi(t)=t^\alpha\sum_{n=0}^\infty a_n t^n$.
In the same way, we can prove the order $\bar{\rho}$ and the type $\bar{\sigma}$ of $\chi(t)$ can be calculate
\begin{align}
\bar{\rho}= \fft{m-2}{2}\,,\qquad |\bar{\sigma}| \bar{\rho}= \sqrt{v_{-m}}\,.
\end{align}
This mean when $r\rightarrow 0$, the wave function $\psi(r)\rightarrow \exp[-\fft{\sqrt{v_{-m}}}{\bar{\rho}}(\fft1r)^{\bar{\rho}}]$. We ignore the solution $\exp[\fft{\sqrt{v_{-m}}}{\bar{\rho}}(\fft1r)^{\bar{\rho}}]$ for boundary condition.
In order to write in a unified form, we let $\bar{\rho}=-\rho$, then we can give a limit
\begin{align}
\lim_{r\rightarrow 0}\psi(r) \propto \exp[\fft{\sqrt{v_{-m}}}{\rho}r^{\rho}]\,,\qquad \rho=\fft{-m+2}{2}\,.
\end{align}
If we replace $-m$ to $m$, this will consistent with our previous results Eq.~(\ref{eq:8}), that we give the case of positive power potentials . Now in order to write in a unified form, we can modify this relation to $|\sigma| \rho= \sqrt{v_{m}}$.

From what have been discussed above, using the node theory, we can ansatz that the wave function have from
\begin{align}
\Psi(r)=f(r)\exp[g(r)]\,.
\end{align}
Where
\begin{align}
f(r)&= \prod_{i=1}^n(r-\alpha_i)\,.\qquad  n=1,2,\cdots,\nn\\
&=1\,.\qquad\qquad \qquad   n=0\,.
\end{align}
And
\begin{align}
g(r)=-\fft{\sqrt{v_{-m}}}{\bar{\rho}}(\fft1r)^{\bar{\rho}}+\cdots+a_{-1}r^{-1}+a_0\ln r+a_1r+\cdots-\fft{\sqrt{v_m}}{\rho}r^{\rho}\,.
\end{align}
Here we choose the minus signature in the terms of $r^{\rho}$ and $(\fft1r)^{\bar{\rho}}$ for the normalization of the wave function.
\section{Harmonic oscillator}
\label{Sec.model1}
Let us consider the one-dimensional Schr\"{o}dinger equation with harmonic oscillator potential $V(r)= A r+B r^2$
\begin{align}
\Psi''(r) = [V(r)-E]\Psi(r)\,.
\end{align}
Make the ansatz $\Psi(r)=f(r)\exp[g(r)]$, let $g(r)= a\ln r+br+cr^2$, and $c= -\frac{\sqrt{B}}{2}$. we have
\begin{align}
\fft{f''+2f'g'}{f}+(g')^2+g'' = V(r)-E\,.
\end{align}
Where the prime denotes the derivation of $r$.

First, we consider the ground state, make $f=1$, then
\begin{align}
 V(r)-E=&\fft{a^2-a}{r^2}+\fft{2ab}{r}+(b^2+4ac+2c)+4bcr\,.
 \label{eq:33}
\end{align}
Compare the coefficients on both sides of Eq.~(\ref{eq:33}), we can obtain the following set of equations
\begin{align}
a^2-a=0\,,\quad 2ab=0\,,\quad-E=b^2-(2a+1)\sqrt{B}\,,\quad -2b\sqrt{B}=A\,.
\end{align}
we can easy solve this equations
\begin{align}
&a=0, \quad b=-\fft{A}{2\sqrt{B}}\,.
\end{align}
According to this choice, they arrive at a constraint on the energy is read as
\begin{align}
E=\sqrt{B}-\fft{A^2}{4B}\,.
\end{align}
The corresponding ground state now be read as
\begin{align}
\Psi_0(r)=N_0 \exp[-\fft{A}{2\sqrt{B}}r-\frac{\sqrt{B}}{2} r^2]\,.
\end{align}
where $N_0=\fft{\sqrt[8]{B}}{\sqrt[4]{\pi}}\exp[-{\fft{A^2}{8B^{3/2}}}]$ is the normalized constant. If set that $A=0$, we found this solution is the ground state of harmonic oscillator.

For the first excited state, we make the ansatz for the first excited state, take $f=r-\alpha_1$, then we have
\begin{align}
 V(r)-E=&\fft{a^2-a}{r^2}+\fft{2ab-2a/\alpha_1}{r}+(b^2+4ac+6c)+4bcr+4c^2r^2\,.
\end{align}
Where the $\alpha_1$ found from the constraint equation
\begin{align}
a+\alpha_1b+2\alpha_1^2c=0.
\end{align}
Compare the coefficients on both sides we can obtain the equations
\begin{align}
&a^2-a=0\,,\quad 2ab-2a/\alpha_1=0\,,\quad-E=b^2+4ac+6c\,,\nonumber\\
&4bc=A\,,\quad 4c^2=B\,.
\end{align}
So we can solve
\begin{align}
&a=0\,,\quad b=-\fft{A}{2\sqrt{B}}\,,\quad c=-\frac{\sqrt{B}}{2}\,,\quad \alpha_1=-\fft{A}{2B}\,.
\end{align}
At the same time, that a restriction on energe give by
\begin{align}
E=3\sqrt{B}-\fft{A^2}{4B}\,.
\end{align}
The corresponding first excited state read as
\begin{align}
\Psi_1(r)=N_1(r-\alpha_1)\exp[br+cr^2]\,.
\end{align}
The parameters $b$, $c$ and $\alpha_1$ are given above, and the normalized constant we can calculate $N_1=\fft{\sqrt{2}{B^{3/8}}}{\sqrt[4]{\pi}}\exp[-{\fft{A^2}{8B^{3/2}}}]$. we also found the first the excited state of harmonic oscillator is equal to $\Psi_1(r)$ under that $A=0$.

Furthermore, with the same spirit, thought make the ansatz $f=(r-\alpha_1)(r-\alpha_2)\cdots$, we can calculate all higher order excited states, and the all solutions are complete.
\section{Hydrogen atom}
\label{Sec.model2}
Let us consider the hydrogen atom, the potential give $V(r)= A/r$,
\begin{align}
\chi''(r) = [V(r)+\frac{l(l+1)}{r^2}-E]\chi(r)\,.
\end{align}
Where the wave function $\Psi(r)=\chi(r)/r$, and $A=e^2$. Ansatz $\chi(r)=f(r)\exp[g(r)]$, let $g(r)= a\ln r+br$, and $b= -{\sqrt{-E}}$. we have
\begin{align}
\fft{f''+2f'g'}{f}+(g')^2+g'' = V(r)-E\,.
\end{align}
Where the prime denotes the derivation of $r$.

First, we consider the ground state $n=0$, make $f=1$, then
\begin{align}
 V(r)-E=&\fft{a^2-a-l^2 - l}{r^2}+\fft{2 a b-A}{r}+b^2\,.
 \label{eq:43}
\end{align}
Compare the coefficients on both sides of Eq.~(\ref{eq:43}), we can easy solve
\begin{align}
&a=-l, \quad b=-\fft{A}{2l}\,,\nn\\
&a=l+1, \quad b=\fft{A}{2(n+l+1)}\,.
\end{align}
Because the bound condition, we will give up the solution about $a=-l$. According to this choice, they arrive at a constraint on the energy is read as
\begin{align}
E=-\fft{A^2}{4(n+l+1)^2}\,.
\end{align}
The corresponding ground state now be read as
\begin{align}
\Psi_0(r)=N_0\fft{\chi_0(r)}{r}=N_0 r^{l}\exp[-{\sqrt{-E}}r]\,.
\end{align}
where $N_0$ is the normalized constant, and for ground state the $l=0$, this solution is the ground state of hydrogen atom.
If $l=1,2,\cdots$, this solution will corresponding to the first(second,and so on) excited state.

For the first excited state $n=1$, we make the ansatz for the first excited state, take $f=r-\alpha_1$, then we have
\begin{align}
 V(r)-E=&\fft{(a^2 + a + 2 \alpha_1 b - l^2 - l)}{r^2}+\fft{2 a b - A + 2 b}{r}+b^2\,.
\end{align}
Where the $\alpha_1$ found from the constraint equation
\begin{align}
a+\alpha_1b=0.
\end{align}
Compare the coefficients on both sides we can solve
\begin{align}
&a=l+1\,,\quad b=\fft{A}{2(n+l+1)}\,,\quad \alpha_1=-\fft{2 (l^2+3 l+2)}{A}\,.
\end{align}
At the same time, that a restriction on energe give by
\begin{align}
E=\fft{A^2}{4(n+l+1)^2}\,.
\end{align}
The corresponding first excited state read as
\begin{align}
\Psi_1(r)=N_1\fft{\chi_1(r)}{r}=N_1(r-\alpha_1)r^{l}\exp[-{\sqrt{-E}}r]\,.
\end{align}
where $N_1$ is normalized constant, and the parameter $\alpha_1$ are given above, The $n$ is the radial quantum number. For first the excited state , because $n=1$, so must be $l=0$, that the wave function $\Psi_1(r)$ corresponding to $R_{20}$.
For the other degenerate state of first the excited state $R_{20}$, corresponding to radial quantum number $n=0$, and angular momentum quantum number $l=1$, give by  $\Psi'_1(r)=N'_1r^{l}\exp[-{\sqrt{-E}}r]$.

Furthermore, with the same spirit, thought make the ansatz $f=(r-\alpha_1)(r-\alpha_2)\cdots$, we can calculate all higher order excited states, and the all solutions are complete.

Similarly, we also can calculate three-dimensional harmonic oscillator in this method, thought ansatz $g(r)= a\ln r+br+cr^2$. This method is not always prefect, in following we will see examples, the cost to solve this Schr\"{o}dinger equation require the parameters of the potential to satisfy certain relations.
\section{Conclusion}
\label{Sec.conclusion}
In this work, we considered solving Schr\"{o}dinger equations with any power series potentials, and using the entire function theory to calculate the order and the type of wave function in all kind of cases. Furthermore, assume wave function satisfaction relation $\psi(r)=f(r)\exp[g(r)]$, with this relation we solving the harmonic oscillator potential and Coulomb potential. 
This method provides us with a new perspective to solve the Schr\"{o}dinger equation.
Although it have some defects, it still give us some enlightenment.
\section*{Acknowledgement}
This work was supported by the Scientific Research Foundation of Guilin University of Technology under grant No. GUTQDJJ2019206.

\appendix

\end{document}